\documentclass{article}

\PassOptionsToPackage{numbers}{natbib}
\usepackage[final]{neurips_2024}

\usepackage[utf8]{inputenc} 
\usepackage[T1]{fontenc}    
\usepackage{hyperref}       
\usepackage{url}            
\usepackage{booktabs}       
\usepackage{amsfonts}       
\usepackage{nicefrac}       
\usepackage{microtype}      
\usepackage{xcolor}         
\usepackage[pdftex]{graphicx}

\setcitestyle{square}

\title{Tuning Music Education: AI-Powered Personalization in Learning Music}

\author{%
  Mayank Sanganeria \\
  Independent Researcher \\
  New York City, USA \\
  \texttt{mayank@ccrma.stanford.edu} \\
  \And
  Rohan Gala \\
  Independent Researcher \\
  Seattle, USA \\
  \texttt{rhngla@gmail.com} \\
}

\begin{document}

\maketitle

\begin{abstract}

Recent AI-driven step-function advances in several longstanding problems in music technology are opening up new avenues to create the next generation of music education tools. Creating personalized, engaging, and effective learning experiences is a continuously evolving challenge in music education. Here we present two case studies using such advances in music technology to address these challenges. In our first case study we showcase an application that uses Automatic Chord Recognition to generate personalized exercises from audio tracks, connecting traditional ear training with real-world musical contexts. In the second case study we prototype adaptive piano method books that use Automatic Music Transcription to generate exercises at different skill levels while retaining a close connection to musical interests. These applications demonstrate how recent AI developments can democratize access to high-quality music education and promote rich interaction with music in the age of generative AI. We hope this work inspires other efforts in the community, aimed at removing barriers to access to high-quality music education and fostering human participation in musical expression.

\end{abstract}

\section{Introduction}

Music holds a unique power to evoke emotions, foster creativity, and build cross-cultural connections. 
However, traditional music education often adopts a one-size-fits-all approach, emphasizing classical repertoires or mainstream genres that may not resonate with the diverse musical preferences of modern students. 
Rigid pedagogical methodology risks alienating learners whose tastes lie beyond the confines of the selected curriculum, hindering engagement and enthusiasm. 
Advances in AI research can be used to offer transformative solutions in this regard -- personalized music education tailored to each student's unique musical identity. 
By curating lessons around an individual's favorite artists, genres, and songs, we can create inclusive environments where students can fulfill their musical potential.

AI music generation has garnered significant attention and made immense progress in recent years~\cite{tahirouglu2023deep, copet2024simple, dhariwal2020jukebox, roberts2018hierarchical}.
We share the opinion that the practice of music is valuable far beyond just the final output \cite{Bown2023Music, kraus2010music}. 
Using generative AI beyond music synthesis, towards analysis of musical concepts from audio data to personalize and support music education remains an under-explored and exciting area for the community.

In the following sections, we first explore the importance of leveraging students' individual musical preferences to enhance motivation and improve learning outcomes. 
We then present two case studies that showcase the potential of AI-driven music technology in education.
One is an ear training application (app) that generates customized exercises based on students' favorite songs. 
The other is an AI-powered piano method book prototype that adapts to students' skill levels and musical interests.

\section{Background}

\subsection*{Need for personalization}

Conventional music curricula tend to prioritize mastery of Western classical music, folk traditions, and works of renowned composers.
While this canonical approach provides a solid foundation, it can fail to captivate students whose musical interests lie in contemporary, non-Western, or niche genres~\cite{hebert2018music}. 
The lack of representation of personal tastes in educational content can disengage learners, limiting their motivation and ability to connect with the material~\cite{woody2021music}. 
As an example, a student drawn into picking up the practice of music through their interest in electronic dance music may find it challenging to engage with a curriculum focused on Baroque-era compositions~\cite{smith2018popular}.

Extensive research highlights the benefits of personalized learning experiences~\cite{shemshack2020systematic}.
Self determination theory suggests that relatedness and autonomy are key components to sustain intrinsic motivation in the pursuit of achievement and performance~\cite{deci2000and}.
Incorporating the student's preferred music and designing ways to interact with it can enhance agency and connection, supporting the educational endeavor~\cite{brown2014make, walkington2014motivating, renwick2012supporting}.

Moreover, the emotional resonance of music is thought to play a pivotal role in its educational efficacy \cite{gusewell2019music}. 
Personalization in music education directly impacts skill acquisition, as engaged students are more likely to practice consistently and persevere through challenges \cite{cogdill2015applying, evans2015self}.

For an individual student, the limitations of traditional music education can be circumvented by working with an exceptional music teacher. Such teachers could incorporate the student's preferred genres with a deep understanding those genres, and with the dedication to create a highly personalized learning experience \cite{rodriguez2017coming, williams2011elephant, vasil2019popular}.
From the teacher's perspective, creating such experiences requires substantial time and effort towards transcribing songs, arranging them to suit the student's current skill level, and designing targeted exercises to develop specific abilities, such as ear training and piano technique. Tools can help ease this burden and enable better teaching of personalized content; even something as simple as automated harmonic analysis of symbolic music, like ChordNamer \footnote{[\url{https://github.com/e7mac/chord_namer}]}.
Given the demands on music teachers, this is uncommon and inaccessible to the majority of music students \cite{kratus2007music}.

\subsection*{AI powered solutions}

We can build digital systems to serve as adaptive learning environments that cater to each student's unique musical interests and learning needs, at scale \cite{williams2007reaching}. 
Through the analysis of audio tracks in each student's listening history, AI could enable creation of dynamic lesson plans and exercises tailored to individual interests as has been explored in language learning settings~\cite{law2024application}. 
Moreover, these adaptive learning systems could monitor student progress, adjusting the content and difficulty of lessons to maintain challenge and interest through the learning process.

The particular technologies we use for our case studies are Automatic Chord Recognition (ACR), beat detection, and Automatic Music Transcription (AMT).

Automatic Chord Recognition (ACR) has evolved considerably from early knowledge-based systems towards data-driven machine learning approaches \cite{pauwels201920}. 
Current tools like ChordAI \footnote{[\url{https://chordai.net}]}, Chordify \footnote{[\url{https://chordify.net}]}, and libraries like crema \footnote{[\url{https://github.com/bmcfee/crema}]}, as well as data-sets like Chord-Annotations \footnote{[\url{https://github.com/tmc323/Chord-Annotations}]} are some of the resources that are available at present. Beat detection is often performed as a step on the way to chord recognition, although dedicated open-source tools such as BeatNet \cite{heydari2021beatnet} also exist.

Automatic Music Transcription (AMT) is the process of converting audio recordings into symbolic musical representations such as sheet music~\cite{benetos2018automatic}.
Researchers have explored options ranging from signal processing techniques to machine learning algorithms \cite{benetos2018automatic} for AMT. 
In recent years, deep learning models have demonstrated remarkable progress in accurately transcribing polyphonic music \cite{gardner2021mt3, sumino2020automatic, kong2021high}, and may soon be application-ready. 
Some of the main tools available at this time are Piano Cover Generation (PiCoGen) \cite{tan2024picogen, tan2024picogen2}, Pop2Piano \cite{choi2023pop2piano}, piano-transcription \footnote{ [\url{https://github.com/bytedance/piano_transcription}]} \cite{kong2021high}, Piano2Notes \footnote{[\url{https://piano2notes.com}]}, and Audio to sheet music converter \footnote{[\url{https://latouchemusicale.com/en/tools/audio-to-sheet-music-converter/}]}.

While it is clear that the underlying music technologies required for music education systems have been researched for some time now, their practical deployment has been limited by accuracy concerns. 
Students must be able to trust the system, which requires the underlying technologies to have very low error rates. 
With AI approaches these technologies are now approaching expert level performance and crossing over to a regime where they can effectively be deployed in educational contexts.

\section{Case Studies}

We now explore how to build such systems through two case studies. 
In the first study we describe our ear training app, which generates custom exercises based on students' favorite songs. 
In the second case study we present an approach to create adaptive and personalized piano method books.

\subsection*{Ear Training App}

\begin{figure}[ht]
    \includegraphics[width=\textwidth]{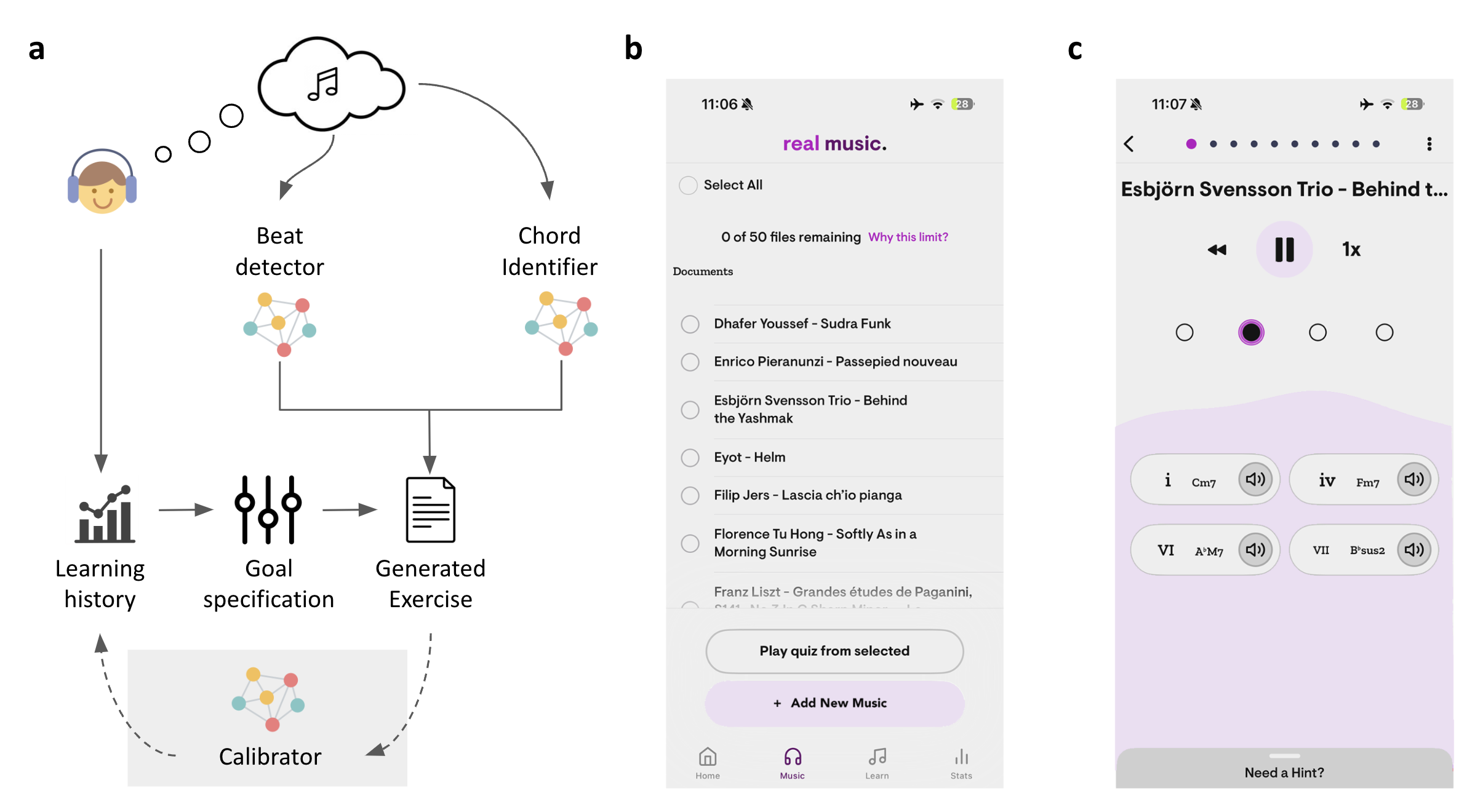}
    \caption{Overview of RealEarTrainer. \textbf{(a)} The student can select custom audio tracks. 
    AI modules detect beats and identify chords within these tracks. 
    The app generates personalized exercises using snippets sourced from the selected audio tracks. 
    Calibrating exercises based on learning history, goal specification, and current performance could be achieved with another AI module. 
    \textbf{(b)} RealEarTrainer interface to select preferred music from a list of available tracks. 
    \textbf{(c)} During the exercise, the app plays a snippet from one of the selected tracks and prompts the student to identify the chords being played. 
    The sound icon on each chord in the options plays a synthesized piano version that can be used by the student to make the harmonic content salient.}
    \label{fig:realeartrainer}
\end{figure}

Traditional ear training tools have relied on the use of piano sounds or the student's primary instrument to facilitate the development of aural skills. 
In real-world musical contexts, students must contend with a myriad of textures, timbres, and parts distributed across multiple instruments. 
To address this challenge, some students resort to training without the aid of dedicated tools, relying solely on their ability to listen to music and reproduce it on their instrument. 
This approach, while valuable, can be time-consuming and may not always provide targeted feedback for improvement.
We've found app-based tools, such as our personal favorite Chet \footnote{[\url{https://chetapp.io}]}, to be effective and helpful not only for building foundational ear training skills but also for bridging the gap between learning environments and real-world musical contexts.

We developed an ear training application that uses beat detection and ACR to analyze student-provided audio files, and generate a personalized ear training curriculum that features snippets and examples drawn directly from the student's preferred music. 
This app is called RealEarTrainer [\url{https://realeartrainer.com}] and is available for download for iOS. Figure \ref{fig:realeartrainer}a-c shows an overview of the app and interface, that we also describe below:

The student begins by selecting their preferred audio tracks within the app.
AI modules identify chords and align them with the beats of each track. This analysis is used to generate a series of tailored exercises.
In each exercise, the student listens to a short snippet sourced from one of their chosen tracks and attempts to identify the chords being played.
The app optionally provides synthesized piano chords aligned with chord changes in the audio snippet. 
This feature is designed to highlight the harmonic content, assisting students until they can confidently identify chords from the original audio alone.

While the core experience of connecting ear training quizzes to the student's preferred music is already functional in the app, we recognize that there is a lot of room in refining how this content is presented to the student. 
In particular, tuning exercise difficulty based on past performance and specific goals (e.g. distinguishing between specific chord families) could be achieved by a calibrator module in future versions.

\subsection*{Personalized piano method book}

Aspiring musicians often find their initial spark of inspiration in a beloved song or piece of music. 
This passion serves as a driving force behind their decision to learn an instrument, towards the goal of being able to play that piece. 
Often, a first step is to pick up a method book for the instrument.

For piano, method books have a rich pedagogical history and intended to teach essential theoretical knowledge and technical skills. 
However, they are limited from a personalization point of view. 
Consequently, students may find themselves with pieces and exercises that, while valuable from an educational perspective, lack the personal resonance and emotional connection that initially drew them to the instrument.
This opens up the prospect of AI-powered piano method books that adapt to each student's musical interests and preferences.
We think personalization can be achieved through two primary approaches.

First, AI can be employed to analyze songs chosen by the student and generate custom arrangements that align with their current skill level and learning objectives. 
This can be achieved through adapting and simplifying the original compositions. 
It could, for instance, remove complex ornamentations, simplify arpeggio patterns and left-hand accompaniment to block chords, among many other simplifications.

Second, AI can be used to create exercises that address technical demands specific to the student's chosen music. 
As opposed to providing a fixed suite of drills, a next generation method book would analyze chord progressions, rhythmic patterns, and melodic elements in the selected songs to construct exercises that directly build the skills necessary to play those specific pieces. 
For example, if a particular song features a complex syncopated rhythm in the left-hand accompaniment, the generated exercises target development of coordination and timing required to execute that specific rhythm accurately. 
These targeted exercises would be presented alongside the simplified arrangement of the song, allowing the student to develop the necessary skills in a focused and purposeful manner.

We demonstrate this concept with a short excerpt from Yann Tiersen's "Comptine D'un Autre Été L'après". We focus on measures 30 - 34 of the piece and prototype an AI system that can simplify the arrangement and generate targeted exercises for a beginner.

\begin{figure}[ht]
    \includegraphics[width=\textwidth]{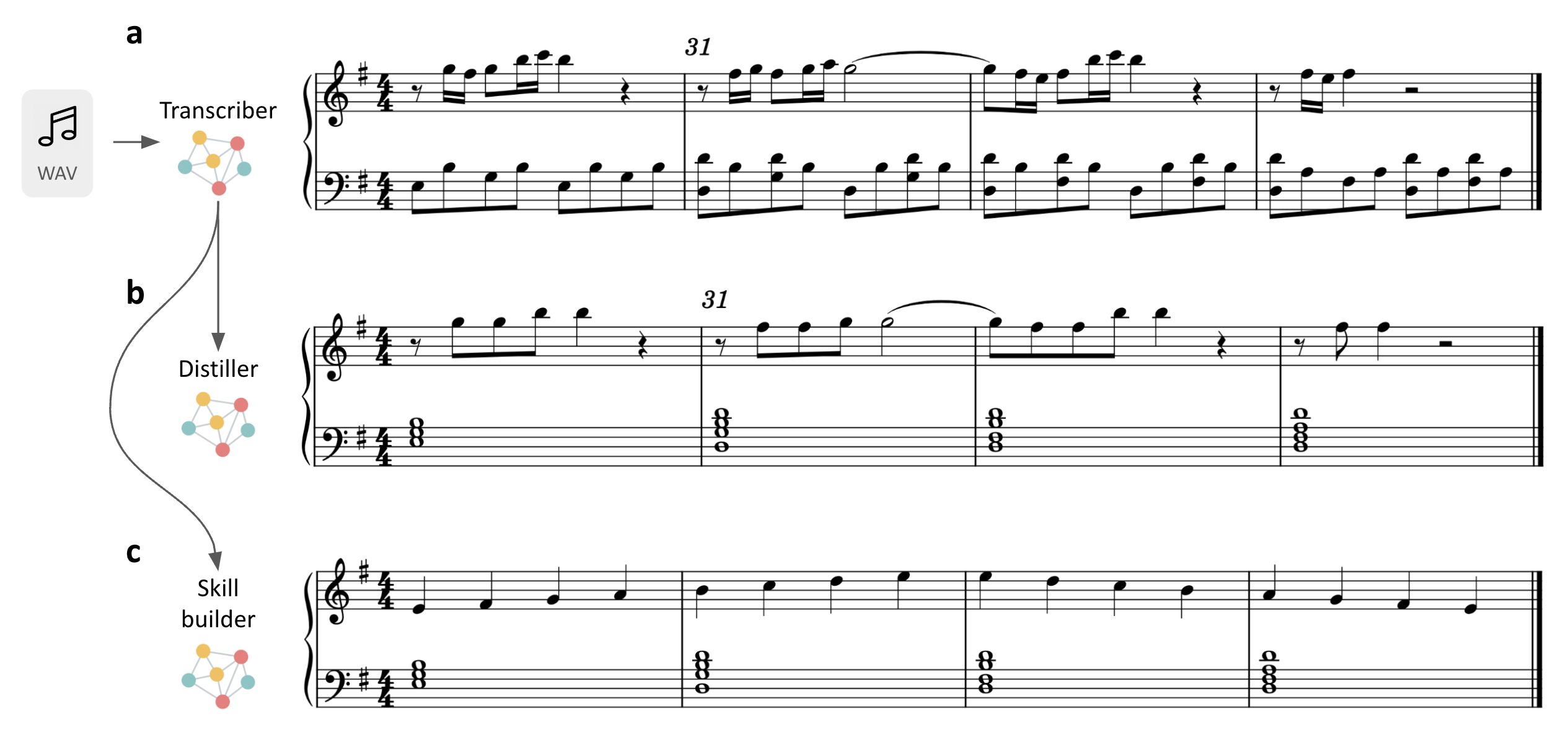}
    \caption{Re-imagined piano method books. \textbf{(a)} We obtain the score for bars 30 - 34 of Comptine D'un Autre Été L'après using Piano2Notes as our transcriber. 
    \textbf{(b)} A mix of procedural and ACR AI modules is used to distill the piece by removing ornamentations and providing block chords that make the core idea easy to follow and play on the piano. 
    \textbf{(c)} Modules to suggest scales over the block chord changes serve to build the related skills while maintaining a connection to the original piece.}
    \label{fig:amelie}
\end{figure}

Our prototype simplifies the left hand part to block chords using the notes present in each measure, and removes the 16\textsuperscript{th} note ornaments in the right hand, Figure \ref{fig:amelie}b. 
This maintains the overall structure and melody of the original piece while being more accessible for a beginner to play. 
This simplified arrangement serves as an achievable milestone on the path to mastering the original composition.

In addition to the simplified arrangement, our prototype generates targeted exercises to help the student develop specific skills required to play this excerpt. 
Figure \ref{fig:amelie}c shows an example exercise focused on learning relevant scales over the chord changes. 
This exercise focuses on playing the scale of the piece, while transitioning between chord changes found in the original excerpt.

Further exercises could target specific skills such as playing the melodic rhythm with the right hand, mastering the rhythm interplay between both hands, and mastering the left-hand accompaniment pattern. 
Breaking down the original piece into manageable components supports the student's progress towards confidently learning the entire piece -- and these can now be created using the original audio file!

\section{Discussion}

AI is poised to enable a paradigm shift in music education, away from rigid curricula prevalent in traditional approaches towards those where personalization take center stage. 
In our opinion, the impact of personalized approaches such as those showcased here for ear training and method books, extends far beyond simple convenience or novelty. 
By directly linking effort and practice to an improved ability to understand, appreciate, and play the music the student loves, such applications can support the learning process for skills that have much broader utility.

Access to skilled teachers is usually limited, and it tends to be expensive as well. 
Through applications like ours, students can receive tailored instruction at a fraction of the cost of private lessons. 
However, we'd like to note that AI-powered music education is not intended to replace human teachers. 
Rather, it can serve as a powerful tool to augment and enhance the work of music educators. 
AI systems can handle the time-consuming tasks of content creation and adaptation, freeing up teachers to focus on higher-level skills such as musical expression, creativity, and collaboration.

While our case studies demonstrate the potential of AI-powered personalization in music education, we acknowledge that comprehensive assessment of these approaches' effectiveness remains as future work. 
Rigorous evaluation through controlled studies comparing traditional and AI-enhanced learning methods, along with longitudinal studies tracking student progress and engagement over time, will be crucial to validate these approaches.

Here we have demonstrated an engaging ear training experience, and a conceptual approach to tailoring method books at various difficulty levels, both enabled by AI. 
These efforts are directed towards lowering barriers, increasing personalization, and we present it with a hope that it inspires other attempts in music education.

\section*{Acknowledgments}
We are grateful to Stochastic Labs for their support and for fostering an innovative environment that has been instrumental in developing RealEarTrainer. Their commitment to exploring the intersection of AI and human creativity aligns with our vision for transforming music education.

We would like to express our sincere gratitude to Vivien Seguy, creator of ChordAI, for his remarkable on Automatic Chord Recognition (ACR), which has been crucial inspiration for these ideas.

\bibliographystyle{unsrtnat}  
\bibliography{references}     

\newpage


\appendix

\section{Supplementary material}

\texttt{simplify.py} is a python script that accepts a MusicXML file as input, replaces the left hand piano part with one consisting only of block chords and the right hand part by modifying 16\textsuperscript{th} note pairs with an 8\textsuperscript{th} note of the first pitch. Also included are the output from the Piano2Notes service, and a MusicXML file corresponding to the 4 bars mentioned in the main text.

\section{Limitations and Future Work}

While this paper explores promising applications of AI in music education, it's important to acknowledge several limitations of our approach and the current state of technology:

\textbf{Accuracy of AI Models:} Despite recent improvements, AI models make mistakes and lead to incorrect feedback or exercises.

\textbf{Limited Scope of Case Studies:} Our case studies focus on ear training and piano education and other areas e.g. composition, music theory, other instruments, require further investigation.

\textbf{Role of Human Music Teachers:} While AI-powered tools offer benefits, they cannot fully replicate the nuanced guidance, emotional support, and real-time adaptability of human music teachers.

\textbf{Cultural Bias:} The AI models used in these applications are likely trained on datasets that may not represent the full diversity of global musical traditions.

\textbf{Limited Assessment:} While we present promising applications, this work does not include comprehensive assessment of their effectiveness. Controlled studies comparing these AI-enhanced approaches with traditional methods are needed. Additionally, while personalization can increase engagement, the long-term effects of these AI-powered approaches on music skill development and retention have not been thoroughly studied. Future work should include:
\begin{itemize}
    \item Comparative studies between traditional and AI-enhanced curricula.
    \item Quantitative metrics for learning outcomes and skill development.
    \item User experience studies with diverse student populations.
    \item Longitudinal research to validate effectiveness over time.
\end{itemize}    
\textbf{Technical Barriers:} The proposed solutions require access to devices capable of running sophisticated AI models, which may not be available to all students, potentially exacerbating educational inequalities.

\textbf{Privacy Concerns:} Analyzing a student's listening history and practice sessions raises important privacy considerations that need to be carefully addressed.


\newpage
\section*{NeurIPS Paper Checklist}

\begin{enumerate}

\item {\bf Claims}
    \item[] Question: Do the main claims made in the abstract and introduction accurately reflect the paper's contributions and scope?
    \item[] Answer: \answerYes{}
    \item[] Justification: The abstract and introduction motivate the need for personalized music instruction, and talk about the case studies presented in the paper.

\item {\bf Limitations}
    \item[] Question: Does the paper discuss the limitations of the work performed by the authors?
    \item[] Answer: \answerYes{}
    \item[] Justification: There is a dedicated section outlining the limitations of the claims in the paper.

\item {\bf Theory Assumptions and Proofs}
    \item[] Question: For each theoretical result, does the paper provide the full set of assumptions and a complete (and correct) proof?
    \item[] Answer: \answerNA{} 
    \item[] Justification: No theoretical results in the paper.

    \item {\bf Experimental Result Reproducibility}
    \item[] Question: Does the paper fully disclose all the information needed to reproduce the main experimental results of the paper to the extent that it affects the main claims and/or conclusions of the paper (regardless of whether the code and data are provided or not)?
    \item[] Answer: \answerYes{} 
    \item[] Justification: We have added outputs from the services used, and the scripts used to manipulate them. As well, links to all the software products, libraries and papers used.

\item {\bf Open access to data and code}
    \item[] Question: Does the paper provide open access to the data and code, with sufficient instructions to faithfully reproduce the main experimental results, as described in supplemental material?
    \item[] Answer: \answerNo{}
    \item[] Justification: The scripts used for the piano exercises have been provided. However, the service used (Piano2Notes) is not open source, so we have provided the output we received from the service. Further, the AI model used in Real Ear Trainer is closed source, but the software is freely available on iOS.

\item {\bf Experimental Setting/Details}
    \item[] Question: Does the paper specify all the training and test details (e.g., data splits, hyperparameters, how they were chosen, type of optimizer, etc.) necessary to understand the results?
    \item[] Answer: \answerNA{}
    \item[] Justification: There are no experiments presented in the paper.

\item {\bf Experiment Statistical Significance}
    \item[] Question: Does the paper report error bars suitably and correctly defined or other appropriate information about the statistical significance of the experiments?
    \item[] Answer: \answerNA{}
    \item[] Justification: There are no experiments presented in the paper.

\item {\bf Experiments Compute Resources}
    \item[] Question: For each experiment, does the paper provide sufficient information on the computer resources (type of compute workers, memory, time of execution) needed to reproduce the experiments?
    \item[] Answer: \answerNA{}
    \item[] Justification: There are no experiments presented in the paper.

\item {\bf Code Of Ethics}
    \item[] Question: Does the research conducted in the paper conform, in every respect, with the NeurIPS Code of Ethics \url{https://neurips.cc/public/EthicsGuidelines}?
    \item[] Answer: \answerYes{} 
    \item[] Justification: Reviewed and confirmed that the paper conforms to the Ethics Guidelines.

\item {\bf Broader Impacts}
    \item[] Question: Does the paper discuss both potential positive societal impacts and negative societal impacts of the work performed?
    \item[] Answer: \answerYes{} 
    \item[] Justification: The paper outlines the positive social impact of making music education more accessible, as well as highlight the potential negative impact on music teachers.

\item {\bf Safeguards}
    \item[] Question: Does the paper describe safeguards that have been put in place for responsible release of data or models that have a high risk for misuse (e.g., pretrained language models, image generators, or scraped datasets)?
    \item[] Answer: \answerNA{} 
    \item[] Justification: No data or models are being released.

\item {\bf Licenses for existing assets}
    \item[] Question: Are the creators or original owners of assets (e.g., code, data, models), used in the paper, properly credited and are the license and terms of use explicitly mentioned and properly respected?
    \item[] Answer: \answerYes{}
    \item[] Justification: Credit has been given, and alternative options have also been provided wherever any assets are mentioned.

\item {\bf New Assets}
    \item[] Question: Are new assets introduced in the paper well documented and is the documentation provided alongside the assets?
    \item[] Answer: \answerNA{} 
    \item[] Justification: No new assets released.

\item {\bf Crowdsourcing and Research with Human Subjects}
    \item[] Question: For crowdsourcing experiments and research with human subjects, does the paper include the full text of instructions given to participants and screenshots, if applicable, as well as details about compensation (if any)?
    \item[] Answer: \answerNA{} 
    \item[] Justification: No crowdsourcing or research with human subjects used in the paper.

\item {\bf Institutional Review Board (IRB) Approvals or Equivalent for Research with Human Subjects}
    \item[] Question: Does the paper describe potential risks incurred by study participants, whether such risks were disclosed to the subjects, and whether Institutional Review Board (IRB) approvals (or an equivalent approval/review based on the requirements of your country or institution) were obtained?
    \item[] Answer: \answerNA{} 
    \item[] Justification: No crowdsourcing or research with human subjects used in the paper.

\end{enumerate}

\end{document}